\documentclass[letterpaper,12pt]{article}   %% LaTeX 2e (preferred)
\usepackage{osajnl2} %% do not use with REVTeX4
\usepackage[draft]{hyperref} %% optional
\usepackage{amsmath}

\begin{document}

\title{Rotation of elliptic optical beams in anisotropic media}

%% For REVTeX it is possible to automate superscript and e-mail callouts with the superscriptaddress option; see REVTeX4 documentation.

\author{Zhixiao Chen and Qi Guo$^{*}$}
\address{Laboratory of Photonic Information Technology, South
China Normal University, Guangzhou 510631, People's Republic of
China}
\address{$^*$Corresponding author: guoq@scnu.edu.cn}

\begin{abstract}We investigate the linear propagation of a paraxial optical beam
in anisotropic media. We start from the eigenmode solution of the
plane wave in the media, then subsequently derive the wave
equation for the beam propagating along a general direction except
the optic axes. The wave equation has a term containing the second
mixed partial derivative which originates from the anisotropy, and
this term can result in the rotation of the beam spot. The
rotation effect is investigated by solving analytically the wave
equation with an initial elliptical Gaussian beam for both
uniaxial and biaxial media. For both media, it is found that there
exists a specific direction, which is dependent on anisotropy of
the media, on the cross-section perpendicular to propagation
direction to determine the rotation of the beam spot. When the
major axis of the elliptical spot of the input beam is parallel to
or perpendicular to the specific direction, the beam spot will not
rotate during propagation, otherwise, it will rotate with the
direction and the velocity determined by input parameters of the
beam.
\end{abstract}

% REPLACE WITH CORRECT OCIS CODES FOR YOUR ARTICLE
                          % NOTE: \ocis{} IS ALIASED TO \pacs{} BUT MUST
                          % FORMAT THE TERMS CORRECTLY FOR EACH JOURNAL

\maketitle %% null function with osajnl.sty

\section{Introduction}
The propagation of light in anisotropic media is still receiving a
good deal of attention from both the experimental and the
theoretical points of view. The subject is of fundamental interest
because of its important role in polarizers, compensators,
amplitude- and phase-modulation devices, also, in nonlinear
optical phenomena which usually take place in anisotropic media.
An accurate electromagnetic description is required for the
evaluation of the field distribution in, for example, designing
polarizers and compensators or in describing the nonlinear
processes like second harmonic generation.

A number of
studies~\cite{fleck,ciattoni1,ciattoni2,ciattoni3,gabriella,ciattoni4,band,marek,dreger,guo,igor,seshadri}
have been devoted to an analysis of optical beam propagation in
anisotropic medium, but most of them are concerned about the
uniaxial cases. For instance, Fleck {\it et al.} \cite{fleck}
derived the paraxial equations for both the ordinary and the
extraordinary components of the field in uniform uniaxial
anisotropic media. The derivation is based on the paraxial
approximation that the electric field corresponding to the beam is
plane polarized with a polarization corresponding to either an
ordinary or an extraordinary wave. Ciattoni {\it et
al.}~\cite{ciattoni1,ciattoni2,ciattoni3,gabriella,ciattoni4}
proposed a theoretical scheme that the exact plane wave angular
spectrum representation is approximated within the paraxial
constraint, obtaining the paraxial expressions for the ordinary
and extraordinary components of a field propagating in uniaxial
media. Trippenbach {\it et al.} \cite{band,marek} considered the
propagation of light pulses in anisotropic and dispersive media,
deriving the equation for the slowly varying envelope of the
electric field. The equation has applications in divers fields. In
their paper, it was used to investigate the propagation of an
optical pulse in uniaxial media, and found that the equation
contains terms that rotate the pulse in time domain. The rotation
effect that originate because of the anisotropy of the medium and
the finite size of the field. The similar effect in spatial domain
can be found in the beam propagation in anisotropic media,
however, it remains undiscussed.

Generally speaking, investigation of optical beam propagation in
biaxial media is more complicated than that in uniaxial media
because of the lower symmetry in the dielectric tensor of the
former. To our knowledge, the first attempt to quantify the
propagation of beam in biaxial media was made by Dreger
\cite{dreger}. In his paper, he derived the optical beam
propagators for biaxial media and specialized them for three cases
of wave-vector direction: general directions, coincident with an
optic axis, and nearly parallel to an optic axis. His derivation
refers to complicated mathematical analysis. In our paper, we will
give another way of deriving the wave equation for paraxial beam
propagating in the general directions.

The paper is organized as follow: In section 2, we use the method
that has been introduced in \cite{born} to obtain the eigenmode
solution in biaxial media. This method is well known, but to our
knowledge, the results for biaxial media are unavailable in the
literatures published so far. In section 3 we derive the equation
for the optical beam in anisotropic media. Previously, Qi Guo and
Sien Chi \cite{guo} have derived the nonlinear paraxial wave
equation for the propagation of an optical beam in anisotropic
media. Because the nonlinear effect are treated as a small
perturbation, the whole derivation is available for the linear
situation if the nonlinear term is not considered. Besides, we
find that there is an improper approximation in the process of the
derivation. In this section, therefore, we adopt their way to
derive the beam equation and make an improvement in the
derivation. In section 4 we illustrate the dynamics of the beam
propagation by using initial elliptical Gaussian beams. Section 5
is a summary.

\section{Eigenmodes in anisotropic media}
We start from the eigenanalysis of optical field in anisotropic
media. From the beginning to Eq. (\ref{derivative}) in this
section, we briefly recall the principles of the methods of how to
obtain the eigenmodes, referring the readers to previous
publications for details.

For a monochromatic optical field $\mathbf{E}(\mathbf
r,t)=\mathbf{E}(\mathbf r)\exp(i\omega t)$ in a magnetically
uniform anisotropic medium, $\mathbf{E}(\mathbf r)$ is described
by the wave equation \cite{fleck}
\begin{equation}\label{maxwell}
\nabla\times[\nabla\times\mathbf{E}(\mathbf{r})]-\frac{\omega^{2}}{c^{2}}\hat{\varepsilon}\cdot\textbf{E}(\mathbf{r})=0,
\end{equation}
where $\hat{\varepsilon}$ is the relative dielectric tensor.
Assuming that Eq. (\ref{maxwell}) has a plane wave solution
\begin{equation}\label{plane-wave-solution}
\mathbf{E}(\mathbf
r)=\mathbf{E}_{0}\exp(-i\mathbf{k}\cdot\mathbf{r}),
\end{equation}where the amplitude
$\mathbf{E}_{0}$ is a constant and $\mathbf{k}$ is the wave
vector, we have the following eigenvalue equation \cite{yariv}
\begin{equation}\label{eigenvalueequation}
(\mathbf{k}\mathbf{k}-k^{2}\hat{I}+\frac{\omega^{2}}{c^{2}}\hat{\varepsilon})\cdot\mathbf{E}_{0}=0,
\end{equation}
where $\hat{I}$ is an unit tensor of rank 2. Moreover, the
dispersion relation
\begin{equation}\label{dispersionequation}
\det(\mathbf{k}\mathbf{k}-k^{2}\hat{I}+\frac{\omega^{2}}{c^{2}}\hat{\varepsilon})=0
\end{equation}can be obtained from the necessary condition for nontrivial solutions
to Eq. (\ref{eigenvalueequation}). On the other hand, the electric
field $\mathbf{E}$ and the electric displacement $\mathbf{D}$ are
related by the constitutive law,
\begin{equation}\label{constitutive law}
\mathbf{D}=\varepsilon_{0}\hat{\varepsilon}\cdot \mathbf{E},
\end{equation}where $\varepsilon_{0}$ is the dielectric constant in
vacuum. The Gauss's law for the plane wave requires that
\begin{equation}
\mathbf{k}\cdot \mathbf{D}=0,
\end{equation}
implying that $\mathbf{k}$ is perpendicular to $\mathbf{D}$. It is
proved \cite{born,hermann} that for every direction of propagation
$\mathbf{k}/|\mathbf{k}|$ for the plane wave in anisotropic media,
there are, in general, two orthogonal polarizations of
$\mathbf{D}$ and, correspondingly, two values of $\mathbf{k}$. The
allowed plane waves are called eigenmodes and their corresponding
polarization directions $\mathbf{D}/|\mathbf{D}|$ are called
eigenvectors \cite{hermann}. As soon as $\mathbf{D}/|\mathbf{D}|$
are found, the polarizations of $\mathbf{E}$ can be obtained from
the inverse function of Eq. (\ref{constitutive law}),
\begin{equation}
\frac{\mathbf{E}}{|\mathbf{E}|}=\frac{\hat{\varepsilon}^{-1}\cdot\frac{\mathbf{D}}{|\mathbf{D}|}}{|\hat{\varepsilon}^{-1}\cdot\frac{\mathbf{D}}{|\mathbf{D}|}|},
\end{equation}
where $\hat{\varepsilon}^{-1}$ is the inverse of the tenser
$\hat{\varepsilon}$. One convenient way of finding
$\mathbf{D}/|\mathbf{D}|$ can be visualized with the aid of the
index ellipsoid \cite{born}, so we are going to solve the
eigenmodes problem in biaxial media by this way.

\begin{figure}[htb]
\centering\includegraphics[width=7cm]{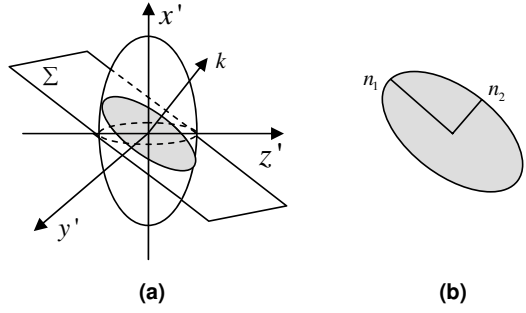}
\caption{\label{ellpisoid-ellipse}(a)Index ellipsoid and (b) the
ellipse that the plane $\Sigma$ cuts in the index ellipsoid.}
\end{figure}

Consider a lossless anisotropic medium, in the principle
coordinate system $(x',y',z')$, characterized by primes, the
dielectric tensor $\hat{\varepsilon}'$ is of a diagonal form,
\begin{equation}\label{dieletrictensor}
\hat{\varepsilon}'= \left(
                                     \begin{array}{ccc}
                                       \varepsilon_{x'} &0&0\\
                                      0& \varepsilon_{y'} & 0 \\
                                      0 & 0& \varepsilon_{z'} \\
                                     \end{array}
                                   \right) ,
\end{equation}
where $ \varepsilon_{x'}$, $\varepsilon_{y'}$ and
$\varepsilon_{z'}$ are real, and
$\varepsilon_{x'}>\varepsilon_{y'}>\varepsilon_{z'}$ is supposed
by convenience. The equation of the index ellipsoid is \cite{born}
\begin{equation}\label{ellipsoid}
\frac{x'^{2}}{\varepsilon_{x'}}+\frac{y'^{2}}{\varepsilon_{y'}}+\frac{z'^{2}}{\varepsilon_{z'}}=1,
\end{equation}
as shown in Fig. \ref{ellpisoid-ellipse}(a). We write the wave
vector $\mathbf{k}$ of the plane wave in the form of
\begin{equation}\label{wavevectork}
\mathbf{k}=\mathbf{e_{x'}}k_{x'}+\mathbf{e_{y'}}k_{y'}+\mathbf{e_{z'}}k_{z'},
\end{equation}
where $\mathbf{e_{x'}}$, $\mathbf{e_{y'}}$ and $\mathbf{e_{z'}}$
are unit vectors along $x'$, $y'$ and $z'$ axes, respectively,
while $k_{x'}=k\cos\alpha$, $k_{y'}=k\cos\beta$ and
$k_{z'}=k\cos\gamma$, with $\cos\alpha$, $\cos\beta$ and
$\cos\gamma$ the direction cosines. The equation of the plane
$\Sigma$ perpendicular to $\mathbf{k}$ through the origin is
\begin{equation}\label{plane}
x'\cos\alpha+y'\cos\beta+z'\cos\gamma=0.
\end{equation}
This plane cuts the ellipsoid in the ellipse (Fig.
\ref{ellpisoid-ellipse} (b)) from which the eigenvectors and the
values of $\mathbf{k}$ can be found \cite{born}: 1, the magnitudes
of the major and minor axes represent the refractive indices of
the eigenmodes which are related to $\mathbf{k}$ by $k=\omega
n/c$, where $c$ is the phase velocity in vacuum; 2, the
eigenvectors are along the corresponding major and minor axes,
respectively.

Since the magnitudes of the major and minor axes, denoted as
$n_{1}$ and $n_{2}$ as shown in Fig. \ref{ellpisoid-ellipse}(b),
are the longest and shortest diameters of the ellipse, they can be
obtained by the Lagrange multiplier method. We can determine
$n_{1}$ and $n_{2}$ by finding the extremum of \cite{born}
\begin{equation}\label{distance}
n^{2}=x'^{2}+y'^{2}+z'^{2}
\end{equation}
subject to the conditions Eq. (\ref{ellipsoid}) and Eq.
(\ref{plane}). According to the Lagrange multiplier method, we
construct the function \cite{born}
\begin{eqnarray}\label{construct}
\nonumber
f=x'^{2}+y'^{2}+z'^{2}+\lambda_{1}\left(\frac{x'^{2}}{\varepsilon_{x'}}+\frac{y'^{2}}{\varepsilon_{y'}}
+\frac{z'^{2}}{\varepsilon_{z'}}-1\right) \\
+\lambda_{2}(x'\cos\alpha+y'\cos\beta+z'\cos\gamma),
\end{eqnarray}
where $\lambda_{1}$ and $\lambda_{2}$ are the two undetermined
multipliers. Now the problem is equivalent to finding the extremum
of $f$ subject to no subsidiary conditions. The necessary
conditions for the extremum of $f$ are that its derivatives with
respect to $x$, $y$ and $z$ should vanish \cite{born}, i.e.
\begin{subequations}\label{derivative}
\begin{equation}\label{fx}
\frac{\partial f}{\partial
x}=2x'+\frac{2x'\lambda_{1}}{\varepsilon_{x'}}+\lambda_{2}\cos\alpha=0
\end{equation}
\begin{equation}\label{fy}
\frac{\partial f}{\partial
y}=2y'+\frac{2y'\lambda_{1}}{\varepsilon_{y'}}+\lambda_{2}\cos\beta=0
\end{equation}
\begin{equation}\label{fz}
\frac{\partial f}{\partial
z}=2z'+\frac{2z'\lambda_{1}}{\varepsilon_{z'}}+\lambda_{2}\cos\gamma=0.
\end{equation}
\end{subequations}
By solving the three equations (\ref{ellipsoid}), (\ref{plane})
and (\ref{derivative}), we can obtain the solutions of $x'$, $y'$,
$z'$, $\lambda_{1}$ and $\lambda_{2}$, then the two extremum of
$n$ ($n$ only takes the positive values) are
\begin{equation}\label{eigenvalue1}
n_{j}=\bigg[\frac{b\pm\sqrt{b^{2}-4a\epsilon_{x'}\epsilon_{y'}\epsilon_{z'}}}{2a}\bigg]^{\frac{1}{2}},
\end{equation}
where $j=1,2$ and
\begin{eqnarray}\label{aandb}
\nonumber
a&=&\epsilon_{x'}\cos^{2}\alpha+\epsilon_{y'}\cos^{2}\beta+\epsilon_{z'}\cos^{2}\gamma
\\
\nonumber
b&=&(\epsilon_{x'}+\epsilon_{y'})\epsilon_{z'}\cos^{2}\gamma+(\epsilon_{x'}+\epsilon_{z'})\epsilon_{y'}\cos^{2}\beta
+(\epsilon_{y'}+\epsilon_{z'})\epsilon_{x'}\cos^{2}\alpha.
\end{eqnarray} The
plus sign is associated with $n_{1}$, the minus sign is associated
with $n_{2}$, respectively. The values of $\mathbf{k}$ are given
by $k_{j}=n_{j}\omega/c$. We call the eigenmode propagating with
the wave vector $\mathbf{k}_{1}$ the mode 1 wave, and that with
$\mathbf{k}_{2}$ the mode 2 wave. With the solutions of
$\lambda_{1}$ and $\lambda_{2}$ inserted into Eq.
(\ref{derivative}), we get the eigenvectors:
\begin{equation}\label{eigenvector1}
\frac{\mathbf{D}_{j}}{|\mathbf{D}_{j}|}=C_{j}\left(\mathbf{e_{x'}}\frac{\epsilon_{x'}\cos\alpha}{\epsilon_{x'}-
n^{2}_{j}}+\mathbf{e_{y'}}\frac{\epsilon_{y'}\cos\beta}{\epsilon_{y'}-n^{2}_{j}}
+\mathbf{e_{z'}}\frac{\epsilon_{z'}\cos\gamma}{\epsilon_{z'}-n^{2}_{j}}\right),
\end{equation}
where
$$
C_{j}=\left[\left( \frac{\epsilon_{x'}\cos\alpha}{\epsilon_{x'}-
n^{2}_{j}} \right)^{2}+\left(
\frac{\epsilon_{y'}\cos\beta}{\epsilon_{y'}-n^{2}_{j}}\right)^{2}+\left(
\frac{\epsilon_{z'}\cos\gamma}{\epsilon_{z'}-n^{2}_{j}}
\right)^{2}\right]^{-\frac{1}{2}}.
$$
It is easy to prove that $\mathbf{D}_{1}\cdot\mathbf{D}_{2}=0$
which means that the two eigenvectors are mutually orthogonal to
each other.

We can see that $n_{1}=n_{2}$ if the polynomial under the radical
sign in Eq. (\ref{eigenvalue1}) is equal to zero,
\begin{eqnarray}
b^{2}-4a\epsilon_{x'}\epsilon_{y'}\epsilon_{z'}=0.
\end{eqnarray}
As $\varepsilon_{x'}>\varepsilon_{y'}>\varepsilon_{z'}$, it can be
rewritten as
\begin{eqnarray}
\nonumber   [\epsilon_{x'}(\epsilon_{y'}-\epsilon_{z'})\cos^{2}\alpha-\epsilon_{z'}(\epsilon_{x'}-\epsilon_{y'})\cos^{2}\gamma]^{2}+\epsilon_{y'}^{2}(\epsilon_{x'}-\epsilon_{z'})^{2}\cos^{4}\beta+\cos^{2}\beta \\
\times\{2\epsilon_{y'}(\epsilon_{x'}-\epsilon_{z'})[\epsilon_{z'}(\epsilon_{x'}-\epsilon_{y'})\cos^{2}\gamma+\epsilon_{x'}(\epsilon_{y'}-\epsilon_{z'})\cos^{2}\alpha]\}=0,
\end{eqnarray}
where the relation that
$\cos^{2}\alpha+\cos^{2}\beta+\cos^{2}\gamma=1$ is used, then we
can obtain
\begin{subequations}\label{optical axes}
\begin{eqnarray}
\cos\beta&=&0,\\
\frac{\cos^{2}\gamma}{\cos^{2}\alpha}&=&\frac{\epsilon_{x'}(\epsilon_{y'}-\epsilon_{z'})}{\epsilon_{z'}(\epsilon_{x'}-\epsilon_{y'})}\equiv\tan^{2}\alpha_{c}.
\end{eqnarray}
\end{subequations} Two directions are specified implicitly by Eq. (\ref{optical
axes}). Along these two directions, we will have $n_{1}=n_{2}$
which shows that the ellipse in Fig. \ref{ellpisoid-ellipse}(b) is
a circle, then the whole formulation breaks down since the
Lagrange multiplier method is unavailable in such case. These two
special directions are known as the optic axes\cite{born}. Eq.
(\ref{optical axes}) shows that they lie in the $x'-z'$ plane, at
angles $\pm\alpha_{c}$ to the $x'$-axis. As the optic axes are the
singularities of the dispersion relations \cite{fleck}, in the
next section about beam propagation, we only discuss the cases
that the optical beam is neither directed along an optic axis nor
so close to an optic axis that any (spatial) spectral component of
the beam is directed along an optic axis. The results for the
uniaxial case can be obtain by assuming that
$\epsilon_{x'}=\epsilon_{y'}$ or $\epsilon_{y'}=\epsilon_{z'}$,
since these results are available in many literatures \cite{born},
we don't present them here.

\section{Wave equation for paraxial beam in anisotropic media}
The wave equation for paraxial beams in anisotropic media had been
derived in reference \cite{guo}, but the derivation has an
improper approximation in the process. In the first subsection, we
will follow the same approach to derive the wave equation and make
an improvement. The wave equation contains coefficients that
account for different propagation behaviors. The expressions of
these coefficients are needed to quantify the beam propagation,
and they have been obtained in the case of uniaxial media
\cite{guo}. In the second subsection, we will discuss the case in
biaxial media and find the expressions of these coefficients.

\subsection{Derivation of wave equation for paraxial beams}
Now, we consider the propagation of an optical beam, which is
formulated by a group of plane waves that belong to only one of
the two mutually orthogonal eigenmodes. The solution of
\ref{maxwell}, therefore, can be expressed as
\begin{equation}\label{field}
\mathbf{E}(\mathbf{r})=\int\int_{-\infty}^{\infty}\mathbf{E}_{0}(\mathbf{k})\exp[i(\mathbf{K}\cdot
\mathbf{R}+k_{z}z)]\rm{d}\mathbf{K},
\end{equation}
where $\mathbf{k}=\mathbf{K}+\mathbf{e_{z}}k_{z}$,
$\mathbf{r}=\mathbf{R}+\mathbf{e_{z}}z$ ($\mathbf{e_{z}}$ is an
unit vector along $z$-axis, the capital letters $\mathbf{K}$ and
$\mathbf{R}$ represent transverse wave vector and transverse
coordinate vector perpendicular to $\mathbf{e_{z}}$, respectively,
$\mathbf{K}=\mathbf{e_{x}}k_{x}+\mathbf{e_{y}}k_{y}$,
$\mathbf{R}=\mathbf{e_{x}}x+\mathbf{e_{y}}y$, with
$\mathbf{e_{x}}$ and $\mathbf{e_{y}}$ defined similar to
$\mathbf{e_{z}}$). In the paraxial beam case, as the wavevectors
of the component waves do not deviate much from the central wave
vector $\mathbf{k}_{0}$, it is convenient to introduce a new
coordinate system $(x, y, z)$, termed propagation coordinate
system hereinafter, so that its $z$-coordinate axis coincides with
$\mathbf{k}_{0}$. In this coordinate system, the input field in
$z=0$ reads
\begin{equation}\label{boundarycondition}
\mathbf{E}_{0}(\mathbf{R},0)=\int\int_{-\infty}^{\infty}\mathbf{E}_{0}(\mathbf{K})\exp(i~\mathbf{K}\cdot\mathbf{R})\rm{d}\mathbf{K}.
\end{equation}

In the spirit of the paraxial approximation, we can assume that
the electric field corresponding to the beam is linearly polarized
with a polarization corresponding to the eigenmode propagating
with the center wave-vector $\mathbf{k}_{0}$ \cite{fleck}.
Concretely, if we write the amplitude of the component waves as
$\mathbf{E}_{0}(\mathbf{K})=\mathbf{p}(\mathbf{K})E_{0}(\mathbf{K})$,
where $\mathbf{p}(\mathbf{K})$ is a unit vector representing the
polarization state, according to the approximation, we have
\begin{equation}\label{talorexpande}
\mathbf{p}(\mathbf{K})\approx \mathbf{p}(0)\equiv\mathbf{p}_{0}.
\end{equation}
This approximation is not the same as that shown in \cite{guo},
where the whole amplitude $\mathbf{E}_{0}(\mathbf{K})$ is
approximated as
$\mathbf{E}_{0}(\mathbf{K})\approx\mathbf{p}_{0}E_{0}(0)$. This
approximation is improper because $E_{0}(0)$ is a constant as soon
as the input field is specified, thus it will make
 Eg. (\ref{boundarycondition}) unable to describe an arbitrary input
field.

under the paraxial condition, $k_{z}$ can be expanded in a Taylor
series at the point $\mathbf{K}=0$,
\begin{equation}\label{talorexpandk}
k_{z}(\mathbf{K})=k_{0}+\nabla_{K}k_{0}\cdot
\mathbf{K}+\frac{1}{2}\nabla_{K}\nabla_{K}k_{0}:\mathbf{K}\mathbf{K}+\cdots,
\end{equation}
with $\nabla_{K}k_{0}=\nabla_{K}k_{z}|_{K=0}$ and so on, and
$\nabla_{K}=\partial/\partial
k_{x}\mathbf{e}_{x}+\partial/\partial k_{y}\mathbf{e}_{y}$. If we
keep terms up to second order in  Eg. (\ref{talorexpandk}) and
make use of  Eg. (\ref{talorexpande}), we can obtain from  Eg.
(\ref{field}),
\begin{equation}\label{pakketform}
\mathbf{E}(\mathbf{r})=\mathbf{p}_{0}A(\mathbf{r})\exp(ik_{0}z) ,
\end{equation}
where
\begin{equation}\label{slowlyvaryingamplitude}
A(\mathbf{r})=\int\int_{-\infty}^{\infty}E_{0}(\mathbf{K})\exp[i\Phi(\mathbf{K},\mathbf{R},z)]\rm{d}\mathbf{K}
\end{equation}
is the slowly varying amplitude, and the phase factor $\Phi$ reads
\begin{eqnarray}
\nonumber
\Phi(\mathbf{K},\mathbf{R},\mathbf{z})=(\nabla_{K}k_{0}\cdot
\mathbf{K}+\frac{1}{2}\nabla_{K}\nabla_{K}k_{0}:\mathbf{K}\mathbf{K})z+\mathbf{K}\cdot\mathbf{R}.
\end{eqnarray}
Introducing  Eg. (\ref{pakketform}) into  Eg. (\ref{maxwell}) with
the help of  Eg. (\ref{eigenvalueequation}) and neglecting the
second derivative $\partial^{2}A/\partial z^{2}$, we can obtain
\cite{guo}
\begin{eqnarray}\label{rawbeamequation}
\nonumber
\int\int_{-\infty}^{\infty}(k_{z}-k_{0})(\mathbf{K}\mathbf{e_{z}}+\mathbf{e_{z}}\mathbf{K})\cdot\mathbf{E}_{0}\exp(i\Phi)\rm{d}\mathbf{K}+(\nabla_{R}\mathbf{e_{z}}+\mathbf{e_{z}}\nabla_{R})\cdot
\mathbf{p}_{0}\frac{\partial A}{\partial \emph{z}} \\
+(\mathbf{e_{z}}\mathbf{e_{z}}-\hat{I})\cdot\mathbf{p}_{0}\left[2ik_{0}\frac{\partial
A}{\partial
z}+\int\int_{-\infty}^{\infty}E_{0}(k_{z}^{2}-k_{0}^{2})\exp(i\Phi)\rm{d}\mathbf{K}\right]=0.\textbf{}
\end{eqnarray}
with $\nabla_{R}=\partial/\partial
x\mathbf{e}_{x}+\partial/\partial y\mathbf{e}_{y}$. It was proved
that the first two terms of  Eg. (\ref{rawbeamequation}) can be
neglected \cite{guo}. From the expansion (\ref{talorexpandk}) we
make the approximation for the factor $(k_{z}^{2}-k_{0}^{2})$ in
the integration,
\begin{equation}\label{k expansion}
k_{z}^{2}-k_{0}^{2}\approx2k_{0}(k_{z}-k_{0})=2k_{0}\nabla_{K}k_{0}\cdot
\mathbf{K}+k_{0}\nabla_{K}\nabla_{K}k_{0}:\mathbf{K}\mathbf{K}.
\end{equation}
Substituting  Eg. (\ref{k expansion}) into Eg.
(\ref{rawbeamequation}), we obtain the beam propagation equation
\cite{guo}
\begin{equation}\label{beam}
i(\frac{\partial A}{\partial z}-\delta_{x}\frac{\partial
A}{\partial x}-\delta_{y}\frac{\partial A}{\partial
y})-\frac{1}{2}\left(\delta_{xx}\frac{\partial^{2}A}{\partial
x^{2}}+2\delta_{xy}\frac{\partial^{2}A}{\partial x\partial
y}+\delta_{yy}\frac{\partial^{2}A}{\partial y^{2}}\right)A=0,
\end{equation}
where $\delta_{i}=\partial_{k_{i}}k_{0}$,
$\delta_{ij}=\partial^{2}_{k_{i},k_{j}}k_{0}$, $i,j=x$ or $y$. Eg.
(\ref{beam}) is a general form of the evolution equation for the
beam in anisotropic media. $\delta_{i}$ and $\delta_{ij}$ are
named $\delta$ coefficients hereafter, and each one has its
physical meaning. The coefficients $\delta_{x}$ and $\delta_{y}$
describe the walk-off of the beam center in $x$ and $y$
directions, respectively. The coefficients $\delta_{xx}$ and
$\delta_{yy}$ are the general Fresnel diffraction coefficients
which are modified by the anisotropy of the medium and are in
general not equal. The coefficient $\delta_{xy}$ involving the
mixed derivative $\partial^{2}A/\partial x\partial y$ vanishes in
isotropic media; it is nonvanishing only if the index of the
refraction depends on the direction of propagation; it can enhance
the diffraction and rotate the cross-section of the beam in the
$x-y$ plane. These effects will be illustrated later.

It should be reminded that the field is determined by Eg.
(\ref{pakketform}) in the lowest-order approximation. Since
$\mathbf{E}(\mathbf{r})$ is proportional to
$\mathbf{p}_{0}A(\mathbf{r})$, the vectorial field is completely
specified when the scalar beam equation Eg. (\ref{beam}) is solved
for $A(\mathbf{r})$. To find the concrete expressions for
different modes, we should first obtain $\mathbf{p}_{0}$ and its
accompanying function $k_{z}(\mathbf{K})$ by making use of the
equations Eg. (\ref{eigenvector1}), Eg. (\ref{dispersionequation})
and Eg. (\ref{eigenvalue1}), and then obtain $k_{0}$ and the
$\delta$ coefficients. In the next section, we will discuss how to
obtain the $\delta$ coefficients in a general biaxial medium, and
how to reduce them in the uniaxial medium case.

\subsection{Expressions of $\delta$ coefficients for biaxial media}
\begin{figure}[htb]
\centering\includegraphics[width=5cm]{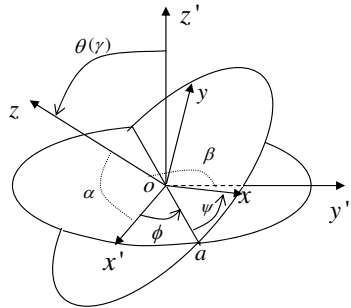}
\caption{\label{euleranglespic}The "$x$-convention" of Euler
angles. The first rotation is by angle $\phi$ about the $z'$-axis,
the second rotation is by an angle $\theta\in[0,\pi]$ about the
former $x$-axis (now $oa$), and the third rotation is by an angle
$\psi$ about the $z$-axis. $\alpha$, $\beta$ and $\gamma$ are the
direction angles of $z$-axis in the principle coordinate system
$(x',y',z')$.}
\end{figure}

Generally, available datas of the dielectric tensor are given in
the principle coordinate system $(x',y',z')$\cite{michael}, in
which $\hat{\varepsilon}'$ is of a diagonal form, but the paraxial
equation  Eg. (\ref{beam}) for the optical beam is obtained in the
propagation coordinate system $(x,y,z)$. To discuss the problem,
therefore, the dielectric tensor should be transformed from the
principal coordinate system to the propagation coordinate system,
in which the dielectric tensor is denoted by $\hat{\varepsilon}$.
To do this, we use the so-called "$x$-convention" definition of
the Euler angles $(\phi,\theta,\psi)$ \cite{weisstein}, as shown
in Fig. \ref{euleranglespic}. In this convention, the coordinate
rotation can be written as
\begin{equation}\label{rotate}
\left(
  \begin{array}{c}
    x \\
    y \\
    z \\
  \end{array}
\right) =\mathbf{T}\left(
                 \begin{array}{c}
                   x' \\
                   y' \\
                   z' \\
                 \end{array}
               \right)
,
\end{equation}
with the rotation matrix $\mathbf{T}$,
\begin{equation}\label{tr}
  \mathbf{T}=\left(
        \begin{array}{ccc}
          \cos\psi\cos\phi-\cos\theta\sin\phi\sin\psi & \cos\psi\sin\phi+\cos\theta\cos\phi\sin\psi & \sin\psi\sin\theta \\
          -\sin\psi\cos\phi-\cos\theta\sin\phi\cos\psi & \cos\theta\cos\phi\cos\psi-\sin\psi\sin\phi & \cos\psi\sin\theta \\
          \sin\theta\sin\phi & -\sin\theta\cos\phi & \cos\theta \\
        \end{array}
      \right).
\end{equation}
Through the angles $\phi$ and $\theta$, we can rotate the $z'$
-coordinate axis to the $z$ -coordinate axis, i.e. the direction
of the central wave vector $\mathbf{k}_{0}$. In the principle
coordinate system, a given propagation direction
$\mathbf{k}_{0}/|\mathbf{k}_{0}|$ is described by the direction
cosines $(\cos\alpha,\cos\beta,\cos\gamma)$, then the values of
Euler angles $\phi$ and $\theta$ can be obtained from the relation
\begin{subequations}\label{eulerangles}
\begin{eqnarray}\label{}
\theta&=&\gamma \\
\tan\phi&=&-\frac{\cos\alpha}{\cos\beta}.
\end{eqnarray}
\end{subequations}
The nutation angle $\psi$ of the final rotation can be arbitrary
chosen since it is used for the orientation of the $x$ and $y$
axes. In general, it is chosen to align the $x$ and $y$
coordinates with the two orthogonal eigenvectors, and these
special values of $\psi$, satisfy
\begin{equation}\label{psi0}
\tan\psi_{0}=\frac{-(\sigma\cot\phi+\tan\phi)}{(1-\sigma)\cos\theta},
\end{equation}
with
$\sigma=(n_{1}^{2}-\varepsilon_{x'})\varepsilon_{y'}[(n_{1}^{2}-\varepsilon_{y'})\varepsilon_{x'}]^{-1}$.
The angle $\psi_{0}$ is to be distinguished from $\psi$ which can
be chosen arbitrarily in defining the propagation coordinate
system. If $\psi=\psi_{0}$, we have
$\mathbf{D}_{1}/|\mathbf{D}_{1}|$ along $x$-axis and
$\mathbf{D}_{2}/|\mathbf{D}_{2}|$ along $y$-axis, respectively. As
a result, $\psi_{0}$ can be used to denote the directions of the
two eigenvectors in propagation coordinate system.

With the coordinates rotation, the dielectric tensor in the
propagation coordinate system becomes
\begin{equation}\label{rotdiele}
\hat{\varepsilon}=\mathbf{T}\hat{\varepsilon}'\mathbf{T^{-1}}=
            \left(
               \begin{array}{ccc}
                 \varepsilon_{11} & \varepsilon_{12} & \varepsilon_{13} \\
               \varepsilon_{12} & \varepsilon_{22} & \varepsilon_{23} \\
               \varepsilon_{13} & \varepsilon_{23} & \varepsilon_{33}
               \end{array}
             \right),
\end{equation}
where $\mathbf{T^{-1}}$ is the inverse matrix of $\mathbf{T}$, and
the components of $\hat{\varepsilon}$ are, respectively,
\begin{eqnarray}\label{epsionalnn}
\nonumber
\varepsilon_{11}&=&(\cos\psi\sin\phi-\cos\theta\sin\psi\sin\phi)^{2}\varepsilon_{x'}
\\
\nonumber &
&+(\cos\theta\cos\phi\sin\psi+\cos\psi\sin\phi)^{2}\varepsilon_{y'}+\sin^{2}\theta\sin^{2}\psi\varepsilon_{z'},
\\
\nonumber
\varepsilon_{12}&=&(-\cos\phi\sin\psi-\cos\theta\cos\psi\sin\phi)(\cos\psi\cos\phi-\cos\theta
\sin\psi\sin\phi)\varepsilon_{x'} \\
\nonumber &
&+(\cos\theta\cos\phi\sin\psi+\cos\psi\sin\phi)(\cos\theta\cos\psi\cos\phi-\sin\psi\sin\phi)\varepsilon_{y'}
\\ \nonumber & &+\cos\psi\sin^{2}\theta\sin\psi\varepsilon_{z'},
\\
\nonumber
\varepsilon_{13}&=&(\cos\psi\cos\phi-\cos\theta\sin\psi\sin\phi)\sin\theta\sin\phi\varepsilon_{x'}
\\
\nonumber &
&-\cos\phi\sin\theta(\cos\theta\cos\phi\sin\psi+\cos\psi\sin\phi)\varepsilon_{y'}+\cos\theta\sin\theta\sin\psi\varepsilon_{z'},
\\
\nonumber
\varepsilon_{22}&=&(-\cos\phi\sin\psi-\cos\theta\cos\psi\sin\phi)^{2}\varepsilon_{x'}
\\
\nonumber &
&+(\cos\theta\cos\psi\cos\phi-\sin\psi\sin\phi)^{2}\varepsilon_{y'}+\cos^{2}\psi\sin^{2}\theta\varepsilon_{z'},
\\
\nonumber
\varepsilon_{23}&=&(\cos\phi\sin\psi-\cos\theta\cos\psi\sin\phi)\sin\theta\sin\phi\varepsilon_{x'}
\\
\nonumber &
&-\cos\phi\sin\theta(\cos\theta\cos\psi\cos\phi-\sin\psi\sin\phi)\varepsilon_{y'}+\cos\theta\cos\psi\sin\theta\varepsilon_{z'},
\\
\nonumber
\varepsilon_{33}&=&\sin^{2}\theta\sin^{2}\phi\varepsilon_{x'}+\cos^{2}\phi\sin^{2}\theta\varepsilon_{y'}+\cos^{2}\theta\varepsilon_{z'}.
\end{eqnarray}
Substituting Eg. (\ref{rotdiele}) into Eg.
(\ref{dispersionequation}), we get the dispersion relationship in
propagation coordinate system. The expressions of $\delta$
coefficients can be obtained from the procedure that introduced in
the last paragraph in section {\it 3.A}. Since the calculation is
complicated, we just present the results here:
\begin{subequations}\label{deltacoefficients}
\begin{equation}\label{gammax}
 \delta_{x}=\frac{\varepsilon_{13}\varepsilon_{22}-\varepsilon_{12}\varepsilon_{23}-\varepsilon_{13}n_{j}^{2}}
{2\varepsilon_{33}n_{j}^{2}-\varepsilon_{11}\varepsilon_{33}-\varepsilon_{22}\varepsilon_{33}+\varepsilon_{13}^{2}+\varepsilon_{23}^{2}},
\end{equation}
\begin{equation}\label{gammay}
 \delta_{y}=\frac{\varepsilon_{11}\varepsilon_{23}-\varepsilon_{12}\varepsilon_{13}-\varepsilon_{23}n_{j}^{2}}
{2\varepsilon_{33}n_{j}^{2}-\varepsilon_{11}\varepsilon_{33}-\varepsilon_{22}\varepsilon_{33}+\varepsilon_{13}^{2}+\varepsilon_{23}^{2}},
\end{equation}
\begin{equation}\label{gammaxx}
 \delta_{xx}=\frac{1}{k_{0}}\left[\frac{-\varepsilon_{12}^{2}-\varepsilon_{13}^{2}+\varepsilon_{11}(\varepsilon_{22}+\varepsilon_{33})-n_{j}^{2}(\varepsilon_{11}+\varepsilon_{33})-4n_{j}^{2}\delta_{x}(\varepsilon_{13}+\varepsilon_{33}\delta_{x})}
{2\varepsilon_{33}n_{j}^{2}-\varepsilon_{11}\varepsilon_{33}-\varepsilon_{22}\varepsilon_{33}+\varepsilon_{13}^{2}+\varepsilon_{23}^{2}}+\delta_{x}^{2}\right],
\end{equation}
\begin{equation}\label{gammayy}
 \delta_{yy}=\frac{1}{k_{0}}\left[\frac{-\varepsilon_{12}^{2}-\varepsilon_{23}^{2}+\varepsilon_{22}(\varepsilon_{11}+\varepsilon_{33})-n_{j}^{2}(\varepsilon_{22}+\varepsilon_{33})-4n_{j}^{2}\delta_{y}(\varepsilon_{23}+\varepsilon_{33}\delta_{y})}
{2\varepsilon_{33}n_{j}^{2}-\varepsilon_{11}\varepsilon_{33}-\varepsilon_{22}\varepsilon_{33}+\varepsilon_{13}^{2}+\varepsilon_{23}^{2}}+\delta_{y}^{2}\right],
\end{equation}
\begin{equation}\label{gammaxy}
 \delta_{xy}=\frac{1}{k_{0}}\left[\frac{\varepsilon_{12}\varepsilon_{33}-\varepsilon_{13}\varepsilon_{23}-n_{j}^{2}\varepsilon_{12}-2n_{j}^{2}\varepsilon_{23}\delta_{x}-2n_{j}^{2}\varepsilon_{13}\delta_{y}-4n_{j}^{2}\varepsilon_{33}\delta_{x}\delta_{y}}
{2\varepsilon_{33}n_{j}^{2}-\varepsilon_{11}\varepsilon_{33}-\varepsilon_{22}\varepsilon_{33}+\varepsilon_{13}^{2}+\varepsilon_{23}^{2}}+\delta_{x}\delta_{y}\right],
\end{equation}
\end{subequations}
where $k_{0}=n_{j}\omega/c$ with $n_{j}$ given by Eg.
(\ref{eigenvalue1}). If $n_{1}$ is substituted, we will obtain a
set of $\delta$ coefficients for the mode 1 beam and that for the
mode 2 beam with $n_{2}$ substituted. In particular,
$\psi=\psi_{0}$ can reduce the $\delta$ coefficients to the same
forms that presented in \cite{dreger}.

Eg. (\ref{beam}) together with the $\delta$ coefficients describes
the paraxial beam propagating in a general direction except along
the optics axes in a general anisotropic medium. In the uniaxial
case, for example, the positive uniaxial crystal with
$\varepsilon_{x'}=\varepsilon_{y'}=n_{o}^{2}$ and
$\varepsilon_{z'}=n_{e}^{2}$ in $\hat{\varepsilon}'$, the mode 1
beam turns out to be the ordinary beam with $n_{1}$ reduced to
$n_{o}$, and the $\delta$ coefficients are reduced to
$\delta_{xx}=\delta_{yy}=-c/(\omega n_{o})$ and zero of the
others, then we obtain the standard paraxial equation for the
ordinary beam,
\begin{equation}\label{standardparaxialequation}
i\frac{\partial A}{\partial z}+\frac{c}{2\omega
n_{o}}\left(\frac{\partial^{2}A}{\partial
x^{2}}+\frac{\partial^{2}A}{\partial y^{2}}\right)A=0.
\end{equation}
On the other hand, the mode 2 beam is the extraordinary beam with
the $\delta$ coefficients reduced to
\begin{subequations}\label{uniaxiadeltacoefficients}
\begin{equation}\label{uniaxialdeltax}
\delta_{x}=\frac{(n_{o}^{2}-n_{e}^{2})\sin\psi\sin2\theta}{2(n_{o}^{2}\sin^{2}\theta+n_{e}^{2}\cos^{2}\theta)},
\end{equation}
\begin{equation}\label{uniaxialdeltay}
\delta_{y}=\frac{(n_{o}^{2}-n_{e}^{2})\cos\psi\sin2\theta}{2(n_{o}^{2}\sin^{2}\theta+n_{e}^{2}\cos^{2}\theta)},
\end{equation}
\begin{equation}\label{uniaxialdeltaxx}
\delta_{xx}=-\frac{1}{k^{(e)}_{0}}\frac{n_{o}^{4}\cos^{2}\psi\sin^{2}\theta+n_{o}^{2}n_{e}^{2}(\sin^{2}\psi\sin^{2}\theta+\cos^{2}\theta)}
{(n_{o}^{2}\sin^{2}\theta+n_{e}^{2}\cos^{2}\theta)^{2}},
\end{equation}
\begin{equation}\label{uniaxialdeltayy}
\delta_{yy}=-\frac{1}{k^{(e)}_{0}}\frac{n_{o}^{4}\sin^{2}\psi\sin^{2}\theta+n_{o}^{2}n_{e}^{2}(\cos^{2}\psi\sin^{2}\theta+\cos^{2}\theta)}
{(n_{o}^{2}\sin^{2}\theta+n_{e}^{2}\cos^{2}\theta)^{2}},
\end{equation}
\begin{equation}\label{uniaxialdeltaxy}
\delta_{xy}=\frac{1}{k^{(e)}_{0}}\frac{n_{o}^{2}(n_{o}^{2}-n_{e}^{2})\sin2\psi\sin^{2}\theta}
{2(n_{o}^{2}\sin^{2}\theta+n_{e}^{2}\cos^{2}\theta)^{2}},
\end{equation}
\end{subequations}
where $k^{(e)}_{0}=n^{(e)}\omega/c$ and
\begin{equation}\label{indexne}
n^{(e)}=\frac{n_{o}n_{e}}{(n_{o}^{2}\sin^{2}\theta+n_{e}^{2}\cos^{2}\theta)^{1/2}}
\end{equation}
is the refractive index with the form expressed in the propagation
coordinate system. Eg. (\ref{indexne}) is obtained by making the
coordinate transformation on $n_{2}$ after it has been reduced
under the uniaxial condition. Here we pay some attention on the
results for the extraordinary beam. The symmetry about the
$z'$-axis of the index ellipsoid makes the propagation direction
only described by $\theta$, so the $\delta$ coefficients are
independent on the angle $\phi$. For the coefficient
$\delta_{xy}$, Eg. (\ref{uniaxialdeltaxy}) shows that once
$\theta$ is specified, its value is modified by the angle $\psi$
and the sine function makes the value periodic. To physically
understand these properties, we recall that the wave vectors of
the extraordinary plane waves belong to an extraordinary ellipsoid
of semi axes $n_{o}\omega/c$, $n_{e}\omega/c$ and $n_{e}\omega/c$,
of which the surface is called the normal surface \cite{hermann}.
For the extraordinary beam, the curved surface of the ellipsoid
near the central wave vector $\mathbf{k}^{(e)}_{0}$ is not
rotational symmetry about $\mathbf{k}^{(e)}_{0}$ if
$\mathbf{k}^{(e)}_{0}$ is not along the optical axis. Therefore,
the mixed derivatives terms arise in the Taylor expansion Eg.
(\ref{talorexpandk}) for a general $\psi$, and its periodic
characteristic is readily understood by recalling that the
orientation of the $x$ and $y$ axes repeat again if $\psi$
increase from zero to $2\pi$. The coefficient $\delta_{xy}$ in the
case of biaxial media has the similar properties but it is not
evident from Eg. (\ref{gammaxy}).

In this uniaxial case, Eg. (\ref{optical axes}) implies that the
optical axis coincides with the $z'$-axis. If we set $\psi=\pi/2$
which lies the $x$-axis in the plane that formed by the optic axis
and the $z$-axis, and substitute (38) into Eg. (\ref{beam}), we
will get the wave equation for the extraordinary beam which is
equal to the linear part of the nonlinear equation obtained by Qi
Guo and Sien Chi (Eq. (42) in \cite{guo}),
\begin{eqnarray}\label{equationinqiguo}
\nonumber & & i\bigg[\frac{\partial A}{\partial
z}-\frac{(n_{o}^{2}-n_{e}^{2})\sin2\theta}{2(n_{o}^{2}\sin^{2}\theta+n_{e}^{2}\cos^{2}\theta)}\frac{\partial
A}{\partial x}\bigg]-\frac{cn_{o}^{2}}{2\omega
n^{(e)}}\bigg[\frac{n_{e}^{2}}
{(n_{o}^{2}\sin^{2}\theta+n_{e}^{2}\cos^{2}\theta)^{2}}\frac{\partial^{2}A}{\partial
x^{2}}\\
& &+\frac{1}
{n_{o}^{2}\sin^{2}\theta+n_{e}^{2}\cos^{2}\theta}\frac{\partial^{2}A}{\partial
y^{2}}\bigg]=0.
\end{eqnarray}
Further more, $\theta=\pi/2$ will reduce Eg.
(\ref{equationinqiguo}) to the form of describing the paraxial
beam propagation orthogonal to the optical axis, which is the same
result given by the equation (8) in reference \cite{ciattoni4}.

\section{Elliptical Gaussian beam propagation}
The properties of paraxial beam propagation in anisotropic media
will be fully elucidated by using a specified input field. In the
first subsection, we will use an initial elliptic Gaussian field
for investigation and obtain the analytic solution. We find that
the elliptic Gaussian beam will rotate during propagation and the
rotation is discussed in detail. In the second subsection, we will
give some examples for illustration.
\subsection{Analytic solution of initial elliptic Gaussian beam}
We discuss the elliptic Gaussian beam propagation in anisotropic
media. Suppose that the input field at $z=0$ is given by
\begin{equation}\label{initialbeam}
\mathbf{E}_{0}(x,y,0)=\mathbf{p}_{0}A_{0}\exp\left(\frac{-x^{2}}
{2w_{x}^{2}}\right) \exp\left(\frac{-y^{2}}{2w_{y}^{2}}\right) ,
\end{equation}
where $w_{x}$ and $w_{y}$ are the initial beam widths in $x $ and
$y$ directions, respectively, and for convenience, $w_{x}>w_{y}$;
$A_{0}$ is the amplitude of the field. With this input field, the
analytic solution of Eg. (\ref{beam}) is \cite{band}
\begin{equation}\label{beamsol}
{A}(x,y,z)=A_{0}\frac{\exp\left\{\frac{-[X+i\delta_{xy}Yz/(w_{y}^{2}-i\delta_{yy}z)]^{2}}
{2[w_{x}^{2}-i\delta_{xx}z+\delta_{xy}^{2}z^{2}/(w_{y}^{2}-i\delta_{yy}z)]}\right\}
\exp\left[\frac{-Y^{2}}{2(w_{y}^{2}-i\delta_{yy}z)}\right]}
{\sqrt{[w_{x}^{2}-i\delta_{xx}z+\delta_{xy}^{2}z^{2}/(w_{y}^{2}-i\delta_{yy}z)](w_{y}^{2}
-i\delta_{yy}z)}} ,
\end{equation}
where $X=x+\delta_{x}z$, $Y=y+\delta_{y}z$, respectively. To
investigate the field evolution, we should first obtain the
modulus of the complex expression Eg. (\ref{beamsol}), which reads
\begin{equation}\label{modulus}
|{A}(x,y,z)|=\frac{\exp\left[-\left(\frac{B_{1}}{2B}X^{2}+\frac{B_{2}}{2B}Y^{2}+\frac{B_{3}}{B}XY\right)\right]}{\sqrt{B}},
\end{equation}
where
\begin{eqnarray}\label{coefficientB}
\nonumber
B_{1}&=&w_{x}^{2}w_{y}^{4}+z^{2}w_{x}^{2}\delta_{yy}^{2}+z^{2}w_{y}^{2}\delta_{xy}^{2},\\
\nonumber
B_{2}&=&w_{x}^{4}w_{y}^{2}+z^{2}w_{y}^{2}\delta_{xx}^{2}+z^{2}w_{x}^{2}\delta_{xy}^{2}
,\\
\nonumber
B_{3}&=&-z^{2}\delta_{xy}(w_{y}^{2}\delta_{xx}+w_{x}^{2}\delta_{yy}),\\
\nonumber
B&=&z^{2}(w_{x}^{2}\delta_{yy}+w_{y}^{2}\delta_{xx})^{2}+[w_{x}^{2}w_{y}^{2}+z^{2}(\delta_{xy}^{2}-\delta_{xx}\delta_{yy})]^{2}.
\end{eqnarray}
The quadratic form of the exponent shows that the field
distribution is a rotated ellipse in $X-Y$ plane if $B_{3}\neq0$.
$B_{3}$ is proportional to $\delta_{xy}$, so $\delta_{xy}=0$ for
no rotation. The rotation angle $\varphi$, defined as the tilt of
the major axis relative to $X$-axis, satisfies
\begin{eqnarray}\label{tan-varph}
\nonumber \tan2\varphi&=&\frac{B_{3}}{B_{1}-B_{2}}\\
&=&\frac{2z^{2}\delta_{xy}(w_{y}^{2}\delta_{xx}+w_{x}^{2}\delta_{yy})}
{w_{x}^{4}w_{y}^{2}-w_{x}^{2}w_{y}^{4}+z^{2}w_{y}^{2}(\delta_{xx}^{2}-\delta_{xy}^{2})-z^{2}w_{x}^{2}(\delta_{yy}^{2}-\delta_{xy}^{2})}.
\end{eqnarray}
A word of conventions is needed here. The rotation angle $\varphi$
determined by Eg. (\ref{tan-varph}) has a range within $\pm\pi/2$,
so it will be the convention of this paper to fix the quantity
unambiguously by defining the arctangent function as having a
range lying in the interval $[-\pi,+\pi]$. On this convention, the
major to semimajor axes ratio $\eta$ of the rotated ellipse is
\begin{equation}\label{widthratio}
\eta=\left(\frac{B_{1}\sin^{2}\varphi+B_{2}\cos^{2}\varphi-\frac{1}{2}B_{3}\sin2\varphi}
{B_{1}\cos^{2}\varphi+B_{2}\sin^{2}\varphi+\frac{1}{2}B_{3}\sin2\varphi}\right)^{1/2}.
\end{equation}

When $\delta_{xy}\neq0$, the rotation velocity, described by
$\partial \varphi/\partial z$, can be obtained from the derivative
of Eg. (\ref{tan-varph}) with respect to $z$,
\begin{equation}\label{rotationvelocity}
\frac{\partial \varphi}{\partial
z}=\frac{2\cos^{2}(2\varphi)zw_{x}^{2}w_{y}^{2}(w_{x}^{2}-w_{y}^{2})\delta_{xy}(w_{y}^{2}\delta_{xx}+w_{x}^{2}\delta_{yy})}
{\left[w_{x}^{4}w_{y}^{2}-w_{x}^{2}w_{y}^{4}+z^{2}w_{y}^{2}(\delta_{xx}^{2}-\delta_{xy}^{2})-z^{2}w_{x}^{2}(\delta_{yy}^{2}-\delta_{xy}^{2})\right]^{2}}.
\end{equation}
Generally, the diffraction coefficients $\delta_{xx}$ and
$\delta_{yy}$ are negatives, making the beam spread in the medium.
As a result, the rotation direction is determined by the
coefficient $\delta_{xy}$ which is directly associated with the
angle $\psi$ while the propagation direction has been specified.
For a chosen $\psi$, if $\delta_{xy}<0$, we have $\partial
\varphi/\partial z>0$, implying that the beam will rotate
counterclockwise. On the contrary, $\delta_{xy}>0$ is
corresponding to the clockwise rotation of the beam. For large
$z$, Eg. (\ref{tan-varph}) is approximate to
\begin{equation}\label{tan-varphlargez}
\tan2\varphi\approx\frac{2\delta_{xy}(\delta_{xx}+\frac{w_{x}^{2}}{w_{y}^{2}}\delta_{yy})}
{\delta_{xx}^{2}-\frac{w_{x}^{2}}{w_{y}^{2}}\delta_{yy}^{2}+(\frac{w_{x}^{2}}{w_{y}^{2}}-1)\delta_{xy}^{2}}.
\end{equation}
Eg. (\ref{tan-varphlargez}) shows that the rotation angle will
approach a value that depends on the ratio of the initial beam
widths $w_{x}/w_{y}$. As a special case, when the input field is a
circular Gaussian beam, i.e. $w_{x}=w_{y}$, it can be proved
directly from Eg. (\ref{beamsol}) that the input field will lose
its circular symmetry during propagation and become elliptical.
This is because the physical properties of the diffractive
spreading (described by $\delta_{xx}$, $\delta_{yy}$,
$\delta_{xy}$) are different in every transverse directions in an
anisotropic media. With $w_{x}=w_{y}$, Eg. (\ref{tan-varph}) gives
\begin{equation}\label{psi1}
\tan2\varphi=\frac{2\delta_{xy}}{\delta_{xx}-\delta_{yy}}.
\end{equation}
The rotation angle is independent on $z$, so the beam will not
rotate but incline by an unvaried angle during propagation.

For the case of $\delta_{xy}=0$, Eg. (\ref{beamsol}) can be
factorized into the product of two one-dimensional Gaussian beams.
In such case, the beam will not rotate but just spread in $x$ and
$y$ directions with the diffraction coefficients $\delta_{xx}$ and
$\delta_{yy}$, respectively.

\begin{figure}[htb]
\centering\includegraphics[width=6cm]{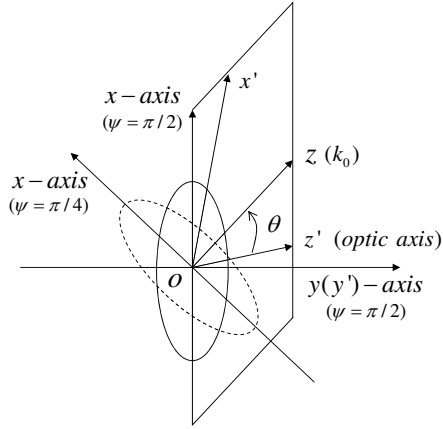}
\caption{\label{uniaxial-different-input}Different values of
$\psi$ correspond to different input fields for the extraordinary
beam in uniaxial crystal. For $\psi=\pi/4$, the input field Eg.
(\ref{initialbeam}) corresponds to the dashed ellipse and this
beam will rotate during propagation; for $\psi=\pi/2$, the input
field Eg. (\ref{initialbeam}) corresponds to the solid ellipse and
this beam will not rotate during propagation}.
\end{figure}

Here we should remind that the angle $\psi$ can be freely chosen
and plays a role of fixing the $x$ and $y$ axes. Since the input
field Eg. (\ref{initialbeam}) is of a regular form in the
propagation coordinate system, different values of $\psi$
correspond to different input fields. If we choose a value of
$\psi$ that makes $\delta_{xy}\neq0$, then the input elliptic
Gaussian beam will rotate during propagation, otherwise, there
will be no rotation effect on the beam. To illuminate the
relationship between $\psi$ and the input field, we still take the
extraordinary beam in the positive uniaxial crystal for example.
The $\delta$ coefficients for the beam are given by Eq.
(\ref{uniaxiadeltacoefficients}). We can obtain from Eg.
(\ref{uniaxialdeltaxy}) that $\delta_{xy}=0$ if $\psi=m\pi/2$,
where $m=0,1,2,3$. If $\psi$ takes one of these values, which
corresponds to lying one of the waists ($w_{x}$ or $w_{y}$) in the
plane that formed by the optic axis and $\mathbf{k}_{0}$, see Fig.
\ref{uniaxial-different-input}, the extraordinary beam will
propagate with no rotation effect (as the case of $\psi=\pi/2$).
Otherwise, the extraordinary beam will rotate during propagation
(as the case of $\psi=\pi/4$). The cases of the biaxial media are
similar except that the values of $\psi$ for $\delta_{xy}=0$ are
different from that in the uniaxial cases.

\subsection{Examples}
\begin{figure}[htb]
\centering\includegraphics[width=6cm]{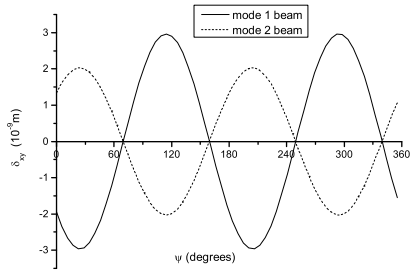}
\caption{\label{deltaxy}$\delta_{xy}$ as a function of $\psi$ with
$\phi=109.50^{\circ}$ and $\theta=6.00^{\circ}$}.
\end{figure}

Here we will use some examples to illustrate the properties of the
propagation that discussed above. For more general, we show the
beams propagation in the biaxial crystal $\rm{KNbO}_{3}$. The
wavelength of the input beams is $\lambda=514\rm{nm}$, and for
this wavelength, the elements of the dielectric tensor in
principle coordinate system are reported as
$\varepsilon_{x'}=5.736$, $\varepsilon_{y'}=5.446$,
$\varepsilon_{z'}=4.893$ \cite{michael} which place the optic axes
at $\alpha_{c}=\pm56.21^{\circ}$. The optical beam is supposed to
propagate along a general direction
$(\alpha,\beta,\gamma)=(84.36^{\circ}, 88^{\circ}, 6^{\circ})$ in
the principle coordinate system. This direction is arbitrary
chosen but should not be too close to the optic axes. It will be
convenient to use this direction in all examples. The refractive
indices of the two eigenmodes are $n_{1}=2.393$ and $n_{2}=2.334$.
According to Eg. (\ref{eulerangles}), the first two Euler angles
are $\phi=109.50^{\circ}$, $\theta=6.00^{\circ}$, and from Eg.
(\ref{psi0}), we have $\psi_{0}=69.94^{\circ}$ . Fig.
\ref{deltaxy} shows the parameter $\delta_{xy}$ versus the angle
$\psi$ in the supposed propagation direction. As shown, there are
four roots for $\delta_{xy}=0$ which are
$69.48^{\circ}+m90^{\circ}$, with $m=0,1,2,3$. To accentuate the
physics of the $\delta_{xy}$ term, we take $\psi=47^{\circ}$ in
the first example to show the propagation of beam with
$\delta_{xy}\neq0$, and then $\psi=69.48^{\circ}$ is taken in the
second example for comparison. The $\delta$ coefficients are
evaluated by (\ref{gammax})-~(\ref{gammaxy}), and the propagation
is computed by using Eg. (\ref{beamsol}) with the propagation
distance normalized by $\xi=zc[\omega(w_{x}^{2}+w_{y}^{2})]^{-1}$.
The results are shown in Fig. \ref{nonzerogammaxy} and Fig.
\ref{zerogammaxy} in which the coordinate frames move with the
centers of the beams, so the walk-off of the beams are not shown
in the figures.

Fig. \ref{nonzerogammaxy} shows the propagation of the two beams
of different modes with three different initial beam widths. Fig.
\ref{nonzerogammaxy}(a) is the case of the initial circular
Gaussian fields. We can see that after the propagation, the
initially circular cross-sections of the two modes both become
elliptical and inclined. The unvaried inclined angles given by Eg.
(\ref{psi1}) are the same, both are equal to $22.48^{\circ}$.
Figures~\ref{nonzerogammaxy}(b) and (c) show the propagation of
the initial elliptical Gaussian beams. The mode 1 beams rotate
counterclockwise and the mode 2 beams rotate clockwise, as
predicted by Eg. (\ref{rotationvelocity}). For $\xi=10$, the
propagation distant is long enough that the rotation angles can be
approximately obtained by Eg. (\ref{tan-varphlargez}): the
rotation angles for the mode 1 beams are
$\varphi\approx76.09^{\circ}$ in (b) and
$\varphi\approx-83.65^{\circ}$ in (c), and for the mode 2 beams,
they are $\varphi\approx-84.52^{\circ}$ in (b) and
$\varphi\approx-86.69^{\circ}$ in (c), respectively. Fig.
\ref{zerogammaxy} shows the cases of $\delta_{xy}=0$. As expected,
there is no rotation or inclination effect on the beams. The beams
just spread in the $X$ and $Y$ directions.

To get a full picture of the beam rotation, we present the details
in the following examples. The parameters are the same as those in
Fig. \ref{nonzerogammaxy} except with $w_{x}=12.5\mu \rm{m}$ and
$w_{y}=10.0\mu \rm{m}$. The results are shown in Fig. \ref{3beams}
which contains four plots with different $\xi$ for each mode. The
mode 1 beam rotates counterclockwise. The rotation angle increases
with increasing $\xi$ and the cross-section keeps elliptical
during propagation. The mode 2 beam has a clockwise rotation. For
$\xi=0.734$, the mode 2 beam has broadened in $Y$ due to Fresnel
diffraction $\delta_{yy}$ and is roughly as broad in $X$ as in
$Y$. Therefore, the tilt of the beam is difficult to see in Fig.
\ref{3beams}(b). As the propagation proceeds, the rotation becomes
evident again, as shown in Fig. \ref{3beams}(c) and (d).

Fig. \ref{2beamscompare}(a) shows the rotation angles versus
$\xi$. It indicates that the rotation angles increase (the mode 1
beam) or decrease (the mode 2 beam) monotonously with the
increasing $\xi$, and for large $\xi$, they both approach certain
values which are given by Eg. (\ref{psi1}), respectively. For the
mode 1 beam, $\varphi\approx66.73^{\circ}$, and
$\varphi\approx-82.69^{\circ}$ for the mode 2 beam. Fig.
\ref{2beamscompare}(b) shows the major to semimajor axes ratio
$\eta$ of the ellipse describing the contour lines of
$|\mathbf{A}(x,y,z)|$. Both of the ratios share the same
characteristics that each of them initially decreases due to the
diffraction and goes through a minimum. Fig.
\ref{2beamscompare}(a) together with (b) shows that before the
$\delta_{xy}$ term makes an obvious rotation on the cross-section
of the beam (for example, $\varphi\approx45^{\circ}$), $\eta$ has
dramatically decreased and approached the minimum value, i.e. the
ellipticity decreases. This is because the diffraction will
exercise the most influence on the evolution of the beam rather
than the rotation effect if the coefficient $\delta_{xy}$ is much
smaller than $\delta_{xx}$ (or $\delta_{yy}$). Although the
rotation will become evident with $\xi$ increasing, the initial
rotation effect is strongly influenced by diffraction and is
barely visible (like the mode 2 beam in Fig. \ref{3beams}(a) and
(b)). As a result, in our examples we choose the spacial widths of
the initial field ($w_{x}$ and $w_{y}$) to be close to each other
so that the diffraction effect is not so strong to influence the
observation of the rotation. The process of beam rotation shown in
Fig. \ref{3beams} may be not so obvious, but we think the examples
are sufficient to elucidate the physics.

An obvious beam rotation can be observed if coefficient
$\delta_{xy}$ is comparable to $\delta_{xx}$ or $\delta_{yy}$ in
magnitude. In order to find a large $\delta_{xy}$, we examine the
ratio $\delta_{xy}/\delta_{xx}$ ($\delta_{yy}$ is on the order of
$\delta_{xx}$). The ratio is a function of the the Euler angles
with the maximum determined by
$(\varepsilon_{x}-\varepsilon_{z})/[2(\varepsilon_{x}\varepsilon_{z})^{1/2}]$
which is associated with the degree of anisotropy. According to
this value, we find that in many natural biaxial crystals like
$\rm{KTP}$, $\rm{LBO}$ or $\rm{LiNbO}_{3}$ \cite{michael},
$\delta_{xy}$ is at least one order less than $\delta_{xx}$. To
our knowledge, natural material with large $\delta_{xy}$ is not
readily available yet. However, in recent scientific progress,
metamaterial technology revolutionized modern optics and photonics
by creating nearly unlimited opportunities for engineering
material parameters. The electric and magnetic properties can be
carefully designed and tuned by changing the geometry, size and
other characteristics of meta-atoms \cite{Werner,litchinitser}.
As a result, material with large coefficient $\delta_{xy}$ might
be realized. For example, supposed that a kind of metamaterial
with a dielectric tenser
$$\hat{\varepsilon}'=\left[%
\begin{array}{ccc}
  6 & 0 & 0 \\
  0 & 4 & 0 \\
  0 & 0 & 2 \\
\end{array}%
\right]$$ has been engineered, according to our theory above, in
the supposed propagation direction, we have
$\delta_{xx}=-8.767\times10^{-8}$,
$\delta_{yy}=-4.107\times10^{-8}$ and
$\delta_{xy}=-2.339\times10^{-8}$ for the mode 1 beam ($\psi$ is
chosen to be equal to$47^{\circ}$). Since $\delta_{xy}$ is
comparable to $\delta_{xx}(\delta_{yy})$, we can expect a
significant effect from $\delta_{xy}$. We present the propagation
of the mode 1 beam in such material in Fig. \ref{beaminmeta}. As
expected, the inclination is obvious.

Through these examples, we can see that the coefficient
$\delta_{xy}$ will play an rotation effect on the beam propagation
if it is nonvanishing. Although here we just present the
propagation of elliptical Gaussian beam, many other simulations
can be made with the beam equation in biaxial media or uniaxial
media, for example, the propagation of higher-order modes of
Hermit-Gaussian beams, we don't present the results here.

\section{Conclusion}
We investigate the propagation of paraxial beam in anisotropic
media. The propagation equation contains terms that account for
the beam walk-off, Fresnel diffraction and beam rotation which is
mainly discussed in this paper. The term responsible for the
rotation originates from the finite beam size and the refractive
index depending on the propagation direction. As a result, this
term vanishes for the wave in isotropic media and for the ordinary
wave in uniaxial media; it is nonvanishing for the extraordinary
wave in uniaxial media and the waves in biaxial media. We
investigate this rotation effect by using an initial elliptical
Gaussian beam in the biaxial media $\rm{KNbO}_{3}$. The analytical
solution shows that the initial elliptical cross-section of the
beam keeps elliptical and rotates clockwise or counterclockwise
during propagation, and for large propagation distance, the
rotation angle will asymptotically approach a certain value which
depends on the ratio of the initial beam widths. It also shows
that an initial circular Gaussian beam will lose its circular
symmetry and become elliptical and inclined. This is because the
anisotropy makes the properties of diffractive spreading different
in every transverse directions. Finally, we point out that the
intensity of the rotation is associated with the degree of
anisotropy. The rotation will be remarkable with high degree of
anisotropy of the media, otherwise, it will be week and may be
covered by diffraction.

\section*{Acknowledgments}
This research was supported by the National Natural Science
Foundation of China (Grant Nos. 10904041 and 10674050),
Specialized Research Fund for the Doctoral Program of Higher
Education (Grant No. 20094407110008).

\clearpage

\section*{List of Figure Captions}

Fig. 1. (a)Index ellipsoid and (b) the ellipse that the plane
$\Sigma$ cuts in the index ellipsoid.

Fig. 2. The "$x$-convention" of Euler angles. The first rotation
is by angle $\phi$ about the $z'$-axis, the second rotation is by
an angle $\theta\in[0,\pi]$ about the former $x$-axis (now $oa$),
and the third rotation is by an angle $\psi$ about the $z$-axis.
$\alpha$, $\beta$ and $\gamma$ are the direction angles of
$z$-axis in the principle coordinate system $(x',y',z')$.

Fig. 3. Different values of $\psi$ correspond to different input
fields for the extraordinary beam in uniaxial crystal. For
$\psi=\pi/4$, the input field Eg. (\ref{initialbeam}) corresponds
to the dashed ellipse and this beam will rotate during
propagation; for $\psi=\pi/2$, the input field Eg.
(\ref{initialbeam}) corresponds to the solid ellipse and this beam
will not rotate during propagation.

Fig. 4. $\delta_{xy}$ as a function of $\psi$ with
$\phi=109.50^{\circ}$ and $\theta=6.00^{\circ}$

Fig. 5. (Color online)Elliptical beams propagation with different
spatial widths of the input fields: (a)$w_{x}:w_{y}=1:1$, (b)
$w_{x}:w_{y}=1.4:1$, (c) $w_{x}:w_{y}=2:1$, where
$w_{y}=10.0\mu\rm{m}$. The contour plots of $|\mathbf{A}(x,y,z)|$
are taken at $\xi=10$. The $\delta$ coefficients are, for the mode
1 beam: $\delta_{x}=1.549\times10^{-2}$,
$\delta_{y}=6.557\times10^{-3}$,
$\delta_{xx}=-3.906\times10^{-8}\rm{m}$,
$\delta_{yy}=-3.486\times10^{-8}\rm{m}$,
$\delta_{xy}=-2.095\times10^{-9}\rm{m}$; for the mode 2 beam:
$\delta_{x}=-1.585\times10^{-3}$, $\delta_{y}=3.745\times10^{-3}$,
$\delta_{xx}=-3.565\times10^{-8}\rm{m}$,
$\delta_{yy}=-3.855\times10^{-8}\rm{m}$,
$\delta_{xy}=1.427\times10^{-9}\rm{m}$, respectively.

\begin{figure}[htb]
\centering
\includegraphics[width=8cm]{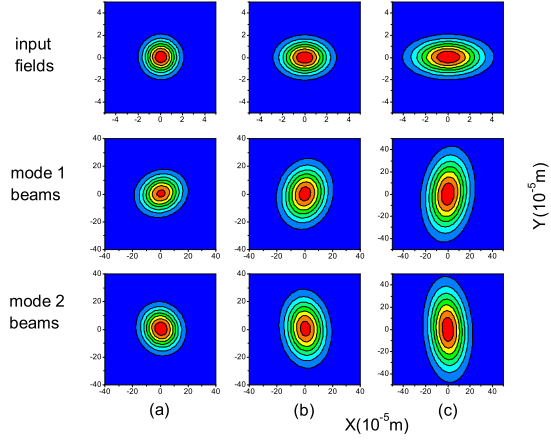}
\caption{\label{nonzerogammaxy}(Color online)Elliptical beams
propagation with different spatial widths of the input fields:
(a)$w_{x}:w_{y}=1:1$, (b) $w_{x}:w_{y}=1.4:1$, (c)
$w_{x}:w_{y}=2:1$, where $w_{y}=10.0\mu\rm{m}$. The contour plots
of $|\mathbf{A}(x,y,z)|$ are taken at $\xi=10$. The $\delta$
coefficients are, for the mode 1 beam:
$\delta_{x}=1.549\times10^{-2}$, $\delta_{y}=6.557\times10^{-3}$,
$\delta_{xx}=-3.906\times10^{-8}\rm{m}$,
$\delta_{yy}=-3.486\times10^{-8}\rm{m}$,
$\delta_{xy}=-2.095\times10^{-9}\rm{m}$; for the mode 2 beam:
$\delta_{x}=-1.585\times10^{-3}$, $\delta_{y}=3.745\times10^{-3}$,
$\delta_{xx}=-3.565\times10^{-8}\rm{m}$,
$\delta_{yy}=-3.855\times10^{-8}\rm{m}$,
$\delta_{xy}=1.427\times10^{-9}\rm{m}$, respectively.}
\end{figure}

\begin{figure}[htb]
\centering\includegraphics[width=8cm]{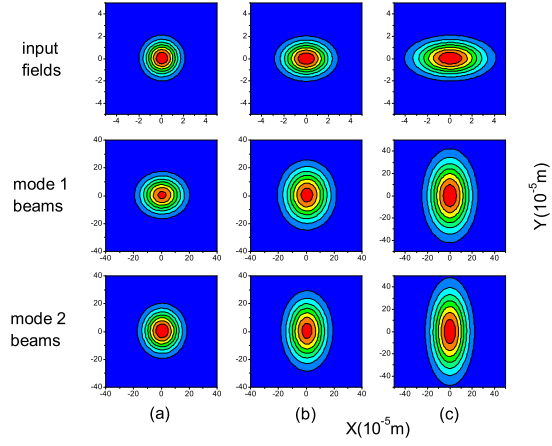}
\caption{\label{zerogammaxy}(Color online)Similar to figure
\ref{nonzerogammaxy} except with $\psi=69.48^{\circ}$. The
$\delta$ coefficients are, for the mode 1 beam:
$\delta_{x}=1.682\times10^{-2}$, $\delta_{y}=1.349\times10^{-4}$,
$\delta_{xx}=-3.992\times10^{-8}\rm{m}$,
$\delta_{yy}=-3.400\times10^{-8}\rm{m}$; for the mode 2 beam:
$\delta_{x}=-4.639\times10^{-5}$, $\delta_{y}=4.067\times10^{-3}$,
$\delta_{xx}=-3.507\times10^{-8}\rm{m}$,
$\delta_{yy}=-3.913\times10^{-8}\rm{m}$, respectively.}
\end{figure}

\begin{figure}[htb]
\centering\includegraphics[width=8cm]{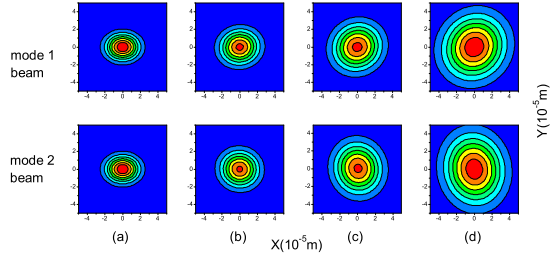}
\caption{\label{3beams}(Color online) Contour of magnitude of the
slowly varying amplitude $|\mathbf{A}(\mathbf{r})|$, for an
initial Gaussian beam with
$\lambda=514\rm{nm}$,$\psi=109.50^{\circ}$, $\theta=6.00^{\circ}$,
and $\psi=47.00^{\circ}$. Spatial widths $w_{x}=12.5\mu \rm{m}$
and $w_{y}=10\mu \rm{m}$ in the frame moving with the center of
the beam for (a)$\xi=0$, (b)$\xi=0.734$, (c)$\xi=1.277$,
(d)$\xi=1.915$.}
\end{figure}

\begin{figure}[htb]
\centering\includegraphics[width=12cm]{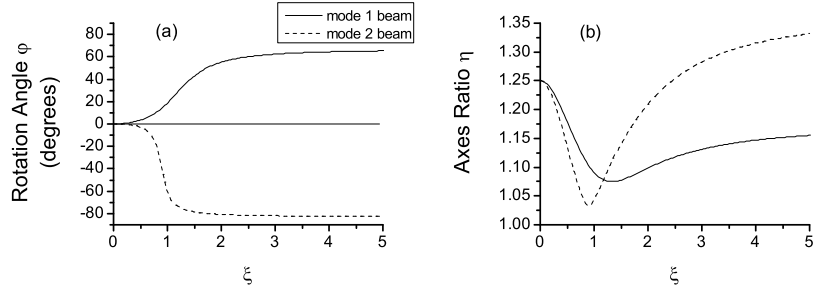}
\caption{\label{2beamscompare}(a)Rotation angle $\varphi$ as a
function of $\xi$. (b) major to semi major axes ratio $\eta$.
Solid curve for the mode 1 beam and dashed curve for the mode 2
beam.}
\end{figure}

\begin{figure}[htb]
\centering\includegraphics[width=9cm]{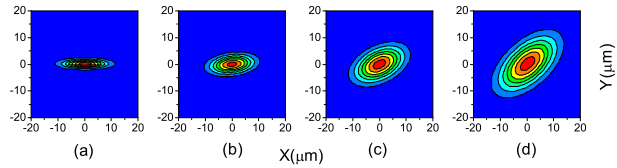}
\caption{\label{beaminmeta}(Color online) Propagation of mode 1
beam in the assumed metamaterial. Spatial widths are $w_{x}=5\mu
\rm{m}$ and $w_{y}=1\mu \rm{m}$ in the frame moving with the
center of the beam for (a)$\xi=0$, (b)$\xi=0.160$, (c)$\xi=0.315$,
(d)$\xi=0.478$.}
\end{figure}

Fig. 6. (Color online)Similar to figure \ref{nonzerogammaxy}
except with $\psi=69.48^{\circ}$. The $\delta$ coefficients are,
for the mode 1 beam: $\delta_{x}=1.682\times10^{-2}$,
$\delta_{y}=1.349\times10^{-4}$,
$\delta_{xx}=-3.992\times10^{-8}\rm{m}$,
$\delta_{yy}=-3.400\times10^{-8}\rm{m}$; for the mode 2 beam:
$\delta_{x}=-4.639\times10^{-5}$, $\delta_{y}=4.067\times10^{-3}$,
$\delta_{xx}=-3.507\times10^{-8}\rm{m}$,
$\delta_{yy}=-3.913\times10^{-8}\rm{m}$, respectively.

Fig. 7. (Color online) Contour of magnitude of the slowly varying
amplitude $|\mathbf{A}(\mathbf{r})|$, for an initial Gaussian beam
with $\lambda=514\rm{nm}$,$\psi=109.50^{\circ}$,
$\theta=6.00^{\circ}$, and $\psi=47.00^{\circ}$. Spatial widths
$w_{x}=12.5\mu \rm{m}$ and $w_{y}=10\mu \rm{m}$ in the frame
moving with the center of the beam for (a)$\xi=0$, (b)$\xi=0.734$,
(c)$\xi=1.277$, (d)$\xi=1.915$.

Fig. 8. (a)Rotation angle $\varphi$ as a function of $\xi$. (b)
major to semi major axes ratio $\eta$. Solid curve for the mode 1
beam and dashed curve for the mode 2 beam.

Fig. 9. (Color online) Propagation of mode 1 beam in the assumed
metamaterial. Spatial widths are $w_{x}=5\mu \rm{m}$ and
$w_{y}=1\mu \rm{m}$ in the frame moving with the center of the
beam for (a)$\xi=0$, (b)$\xi=0.160$, (c)$\xi=0.315$,
(d)$\xi=0.478$.

%\noindent Fig. 2. ...
%\noindent Fig. 3. ...

%\listoffigures

\clearpage

%% sample sizing command; other sizing commands (and graphics packages) may be used as well

\end{document}